\documentclass[12pt]{article}
\usepackage{epsfig,float}
\newcommand{\nn}{\nonumber}
\newcommand{\bea}{\begin{eqnarray}}
\newcommand{\eea}{\end{eqnarray}}
\def \branch{{\cal B}}
\def \beq{\begin{equation}}
\def \eeq{\end{equation}}
\def \branch{{\cal B}}
\def \Im{{\hbox{Im}}\,}
\def \Re{{\hbox{Re}}\,}
\def \gev{{\hbox{GeV}}}
\def \mev{{\hbox{MeV}}}
\def \tev{{\hbox{TeV}}}
\def \cl#1{{#1\%\ \mathrm{C.L.}}}
\def \eq#1{Eq.~(\ref{#1})}

\def \fig#1{Fig.~\ref{#1}}

\def \nn{\nonumber}
\def \rf{Ref.~\cite}

\def \bit{\begin{itemize}}
\def \eit{\end{itemize}}
\def \lu{\lambda_u}
\def \a{\alpha}
\def \b{\beta}
\def \D{\Delta}
\def \g{\gamma}
\def \G{\Gamma}
\def \d{\delta}

\def \l{\lambda}

\def \m{\mu}

\def \p{\pi}
\def \r{\rho}

\def \z{\zeta}
\def \et{\eta}

\begin{document}

\begin{titlepage}

\begin{flushright}
DESY 02-089
\end{flushright}

\begin{center}
\bf \large Implications of $B \to \rho \gamma$ measurements in
 the \\ Standard Model and Supersymmetric Theories
\end{center}

\bigskip

\begin{center}
\large
A. Ali\footnote{E-mail address: ali@mail.desy.de} ~and
E. Lunghi\footnote{E-mail address: lunghi@mail.desy.de}
\end{center}

\begin{center}
Deutsches Elektronen Synchrotron, DESY, \\
Notkestrasse 85, D-22607 Hamburg, Germany
\end{center}

\bigskip

\begin{abstract}
  We study the implications of the recently improved upper limits on
  the branching ratios for the decays $B \to \rho \gamma$, expressed
  as $R(\rho \gamma/K^* \gamma)\equiv {\cal B}(B \to \rho
  \gamma)/{\cal B}(B \to K^* \gamma) <0.047$. We work out the
  constraints that the current bound on $R(\rho \gamma/K^* \gamma)$
  implies on the parameters of the quark mixing matrix in the standard
  model (SM). Using the present profile of the unitarity triangle, we
  predict this ratio to be $R(\rho\gamma/K^*\gamma)=0.023 \pm 0.012$.
  We also work out the correlations involving $R(\rho \gamma/K^*
  \gamma)$, the isospin-violating ratio $\Delta (\rho \gamma)$, and
  the direct CP-violating asymmetry $A_{\rm CP}(\rho \gamma)$ in $B
  \to \rho \gamma$ decays in the SM, in the minimal supersymmetric
  extension of the SM (MSSM), and in an extension of the MSSM
  involving an additional flavor-changing structure in $b \to d$
  transitions.
\end{abstract}

\end{titlepage}

\section{Introduction}
Recently, the BABAR collaboration has reported a significant
improvement on the upper limits of the branching ratios for the
decays $B^0(\bar B^0) \to \rho^0\gamma$ and $B^\pm \to \rho^\pm
\gamma$. Averaged over the charge conjugated modes, the current
$\cl{90}$ upper limits are~\cite{babar:convery}:
\bea
{\cal B}(B^0 \to \rho^0 \gamma) &<& 1.4 \times 10^{-6} \; , \\
{\cal B}(B^\pm \to \rho^\pm \gamma) &<& 2.3 \times 10^{-6} \; , \\
{\cal B}(B^0 \to \omega \gamma) &<& 1.2 \times 10^{-6} \; .
\eea
They have been combined,
using isospin weights for $B \to \rho \gamma$ decays and assuming
${\cal B}(B^0 \to \omega \gamma)={\cal B}(B^0 \to \rho^0 \gamma)$,
to yield the improved upper limit
\beq
{\cal B}(B \to \rho \gamma) < 1.9 \times 10^{-6}\; .
\eeq
 The current measurements of the branching
ratios for $B \to K^* \gamma$ decays by BABAR~\cite{babar:grauges},
\bea
{\cal B}(B^0 \to K^{*0} \gamma)&=&(4.23 \pm
0.40 \pm 0.22) \times 10^{-5} \; , \\
{\cal B}(B^+ \to K^{*+} \gamma) &=&(3.83 \pm 0.62 \pm 0.22) \times 10^{-5}\; ,
\eea
are then used to
set a $\cl{90}$ upper limit on the ratio of the branching ratios~\cite{babar:convery}
\bea
& & R(\rho \gamma/K^*\gamma) \equiv  \frac{{\cal B}(B \to \rho \gamma)}{{\cal B}(B
\to K^* \gamma)} < 0.047 \; .
\eea
This bound is typically a factor 2 away from the SM
estimates~\cite{bdgAP}, which we quantify more precisely in this
letter. In beyond-the-SM scenarios, this bound provides a highly
significant constraint on the relative strengths of the $b \to d
\gamma$ and $b \to s \gamma$ transitions.

The impact of the measurement of $R(\rho \gamma/K^*\gamma)$ on the
parameters of the quark mixing matrix (henceforth called the
Cabibbo-Kobayashi-Maskawa CKM matrix) has been long anticipated (see,
for example, \cite{Ali:vd}). This quantity measures essentially the
CKM matrix element ratio $\vert V_{td}\vert^2/\vert V_{ts}\vert^2$ in
the SM. However, one expects significant long-distance contributions
in $R(\rho \gamma/K^*\gamma)$ entering in the decay $B \to \rho
\gamma$.  They are dominated by the annihilation diagrams $b\bar{u}
\to d\bar{u} \gamma$ in the decays $B^- \to \rho^- \gamma$
\cite{Ali:1995uy,Khodjamirian:1995uc,Grinstein:2000pc,Beyer:2001zn},
which depend on the CKM matrix elements $V_{ub}V_{ud}^*$. The
corresponding annihilation contribution in the decays $B^0 \to \rho^0
\gamma$ (and its charge conjugate) is parametrically suppressed due to
the electric charge of the spectator quark in $B^0$ and the
unfavorable color factors. QCD corrections to the decay widths for $B
\to \rho \gamma$ also introduce a dependence on $V_{ub}V_{ud}^*$ in
both the charged and neutral $B$-meson decays. As the relevant CKM
matrix element ratio $\lambda_u \equiv V_{ub}V_{ud}^*/V_{tb}V_{td}^*$
is of $O(1)$, these modifications are important and have to be taken
into account in the analysis of $R(\rho \gamma/K^*\gamma)$ and other
observables in $B\to \rho \g$ decays.

Recently, the $O(\alpha_s)$ corrections in the decay widths for $B \to
V \gamma$ $(V=K^*,\rho)$ have been calculated in the context of a QCD
factorization framework~\cite{Beneke:1999br}, taking into account the
explicit $O(\alpha_s)$ and $1/M_B$ corrections to the penguin
amplitudes \cite{bdgAP,Bosch:2001gv,Beneke:2001at}. Using the
theoretical results at hand, we analyze the impact of the current
upper limit $R(\rho \gamma/K^* \gamma)< 0.047$ in the context of the
SM, where it yields constraints on the CKM parameters $\bar{\rho}$ and
$\bar{\eta}$ \cite{bdgAP}, and in some popular extensions of the SM,
such as the minimal flavor violating minimal supersymmetric standard
model (MFV-MSSM)~\cite{mfvCDGG}, and in an Extended--MFV-MSSM scenario
(EMFV) ~\cite{emfvAL}, having a non-CKM flavor-changing structure
involving the $b \to d$ transition.  We also present the correlations
involving $R(\rho \gamma/K^* \gamma)$, the isospin-violating ratio
$\Delta (\rho \gamma)$, and the direct CP-violating asymmetry $A_{\rm
CP}(\rho \gamma)$ in $B \to \rho \gamma$ decays, in the three models
just mentioned. Precise measurements of these correlations would
provide a strong discrimination among the competing models.

\section{Observables}
The effective Hamiltonian for the radiative decays
$B \to \rho \gamma$ (equivalently $b \to d \gamma$ decay) can be seen
for the SM in \cite{bdgAP}. We shall invoke this effective Hamiltonian
also for the MFV-MSSM and the EMFV cases, which differ from the SM in
the Wilson coefficients (WC's), in particular in the effective WC's
for the magnetic moment operator, $C_7^s$ (for $b \to s \gamma)$ and
$C_7^d$ (for $b \to d \gamma)$. While this is certainly not the most
general operator basis, it is a sufficient basis to illustrate the
beyond-the-SM effects that may arise in these decays.  Restricting
ourselves to this basis, we first present the $O(\alpha_s)$-corrected
expressions for the observables in the $B\to \r\g$ decays, worked out
in the SM in \cite{bdgAP,Bosch:2001gv}, but now generalized to the
case of complex Wilson coefficients.
\begin{table}[!t]
\begin{center}
\begin{tabular}{|l|l|} \hline
 $\z = 0.76 \pm 0.10$ & $L^u_R = -0.095 \pm 0.022 $ \cr
 $A^{(1)K^*} = -0.113 - i 0.043$ & $A^{(1)t} = -0.114 - i 0.045$ \cr
 $A^u = -0.0181 + i 0.0211$ & \cr \hline
 $\eta_{tt} = 0.57$ & $\eta_{cc} = 1.38 \pm 0.53$ \cr
 $\eta_{tc} = 0.47 \pm 0.04$ & $\hat B_K = 0.86 \pm 0.15$ \cr
 $\eta_B = 0.55$ & $f_{B_d} \sqrt{\hat B_{B_d}} = 235 \pm 33^{+0}_{-24}\; \mev$ \cr
 $\xi_s= 1.18 \pm 0.04^{+0.12}_{-0}$  & \cr \hline
 $\l = 0.221 \pm 0.002$ & $|V_{ub}/V_{cb}| = 0.097 \pm 0.010$ \cr
 $\epsilon_K = (2.271 \pm 0.017) \; 10^{-3}$ & $\D M_{B_d} = 0.503 \pm
0.006 \; {\rm ps}^{-1}$ \cr
 $a_{\psi K_s} = 0.734 \pm 0.054$ & $\D M_{B_s} \geq 14.4 \; {\rm ps}^{-1}
\; (\cl{95})$   \cr \hline
\end{tabular}
\caption{\it Theoretical parameters and measurements used in $B\to
\r\g$ observables and in the CKM unitarity fits.}
\label{inputs}
\end{center}
\end{table}

Here and in the following we will always consider quantities averaged
over the charge conjugated modes (with the obvious exception of the
CP asymmetries). Starting from the decay widths
$\Gamma (B^+ \to V^+ \g)$, $\Gamma (B^- \to V^- \g)$,
$\Gamma (B^0 \to V^0 \g)$ and $\Gamma (\bar B^0 \to \bar V^0 \g)$ (with
$V=\rho,\; K^*$), we construct
\bea
\Gamma^\pm (B \to V \g) = \frac{\Gamma (B^+ \to V^+ \g)+\Gamma (B^- \to V^- \g)}{2} \; , \\
\Gamma^0   (B \to V \g) = \frac{\Gamma (B^0 \to V^0 \g)+\Gamma (\bar B^0 \to \bar V^0 \g)}{2} \; .
\eea
We will define the various observables in terms of these quantities.
Note that, up to the NLO approximation, this procedure is equivalent to
defining two
distinct observables for the charge conjugate modes and {\em then}
performing the average.
It is preferable to use the above definitions since they involve quantities
(the CP averaged decay widths) that are much easier to measure than the
widths of the individual channels, which would require tagging the
$B$-meson.

The expression for the ratios $R (\rho \gamma/K^* \gamma)$ is~\cite{bdgAP}
\bea
R^\pm (\rho \gamma/K^ \gamma) &=&  \left| V_{td} \over V_{ts} \right|^2
 {(M_B^2 - M_\r^2)^3 \over (M_B^2 - M_{K^*}^2)^3 } \z^2
(1 + \D R^\pm) \; , \\
R^0 (\rho \gamma/K^* \gamma) &=& {1\over 2} \left| V_{td} \over V_{ts} \right|^2
 {(M_B^2 - M_\r^2)^3 \over (M_B^2 - M_{K^*}^2)^3 } \z^2
(1 + \D R^0) \; ,
\label{rapp}
\eea
where $\z=\xi_{\perp}^{\rho}(0)/\xi_{\perp}^{K^*}(0)$, with
$\xi_{\perp}^{\rho}(0) (\xi_{\perp}^{K^*}(0))$ being the form factors
(at $q^2=0$) in the effective heavy quark theory for the decays $B \to
\rho (K^*)\gamma$. Noting that in the SU(3) limit one has $\z=1$,
some representative estimates of the SU(3)-breaking and the resulting
values of $\z$ are: $\z=0.76 \pm 0.06$ from the light-cone QCD
sum rules~\cite{Ali:1995uy}; a theoretically improved estimate in the
same approach yields \cite{Ball:1998kk}: $\z=0.75 \pm 0.07$;
$\z =0.88 \pm 0.02$  using hybrid QCD sum
rules~\cite{Narison:1994kr}, and $\z=0.69 \pm 10\%$ in the quark
model~\cite{Melikhov:2000yu}. Except for the hybrid QCD
sum rules, all other approaches yield a significant SU(3)-breaking in the
magnetic moment form factors. In the light-cone QCD sum rule approach,
this is anticipated due to the appreciable differences in the wave
functions of the $K^*$ and $\rho$-mesons. To reflect the current
dispersion in the theoretical estimates of $\z$, we take its value as
$\z=0.76 \pm 0.10$, given in  Table~\ref{inputs}. We stress that the error
($\pm 0.10)$ is not on $\z=0.76$, but rather on the deviation of $\z$ from
its SU(3) limit, i.e.,  $1-\z=0.24$, and amounts to an error of $\pm 42\%$
on the SU(3)-breaking in the ratio of form factors in radiative decays.
As this is the dominant theoretical error on the ratios $R^\pm (\rho
\gamma/K^ \gamma)$ and  $R^0 (\rho \gamma/K^* \gamma)$, it is imperative
to reduce it. A lattice-QCD based estimate of the form factors, and hence
$\z$, is highly desirable.

 The quantity $(1 +\Delta R)$ entails
the explicit $O(\alpha_s)$ corrections, encoded through the functions
$A_{\rm R}^{(1)K^*}$, $A_{\rm R}^{(1) t}$ and $A_{\rm R}^{u}$, and the
long-distance contribution $L_{\rm R}^{u}$. For the decays $B^\pm \to
\rho^\pm \gamma$ and $B^\pm \to K^{*\pm} \gamma$, this can be written
after charge conjugated averaging as
\beq
1 + \D R^\pm = \left| C_7^d + \lu L^u_R \over C_7^s \right|^2
           \left( 1 - 2 A^{(1)K^*}_R  {\Re C_7^s \over |C_7^s|^2} \right)
+ {2 \; \Re \left[ (C_7^d + \lu L^u_R)
 (A^{(1)t}_R + \lu^* A^u_R) \right]\over |C_7^s|^2} \; .
\label{dr}
\eeq
The definitions of the quantities $A^{(1)K^*}$, $A^{(1)t}$, $A^u$ and
$L^u_R$ can be seen in \cite{bdgAP}.  Their default values
are summarized in Table~\ref{inputs}, where we have
also specified the theoretical errors in the more sensitive of
these parameter $L^u_R$. The quantity $1 + \D R^0$ is obtained from
Eq.~(\ref{dr})
in the limit $L^u_R = 0$.

The isospin breaking ratio is given by
\bea
\D (\r\g) &=& {\Gamma^\pm (B \to \r \g) \over 2 \; \Gamma^0 (B  \to \r \g)} -1 \\
&=&
\left| C_7^d + \lu L^u_R \over C_7^d \right|^2
              \left( 1 - {2 \Re C_7^d ( A^{(1)t}_R + \lu^* A^u_R)\over |C_7^d|^2} \right) \nn \\
 & & +{2 \; \Re \left[ (C_7^d + \lu L^u_R) (A^{(1)t}_R + \lu^* A^u_R) \right]\over |C_7^d|^2}-1\; ,
\label{dnlo}
\eea
and the CP asymmetry in the charged modes is
\bea
A^\pm_{CP} (\r\g) &=& \frac{\branch (B^- \to \r^- \g) -
\branch (B^+ \to \r^+ \g) }{\branch (B^- \to \r^- \g) + \branch (B^+
\to \r^+ \g) } \\
 &=& - {2  \Im \left[ (C_7^d + \lu L^u_R) (A^{(1)t}_I + \lu^* A^u_I) \right]
                   \over |C_7^d + \lu L^u_R|^2} \; .
\label{acp}
\eea
The CP asymmetry in the neutral modes
\bea
A^0_{CP} (\r\g) &=& \frac{\branch (B^0 \to \r^0 \g) -
\branch (\bar B^0 \to \r^0 \g) }{\branch (B^0 \to \r^0 \g) + \branch (\bar B^0
\to \r^0 \g) }
\eea
is obtained from Eq.~(\ref{acp}) in the limit $L^u_R = 0$.

 \begin{figure}[!t]
 \centerline{\mbox{\epsfxsize=\linewidth \hbox{\epsfbox{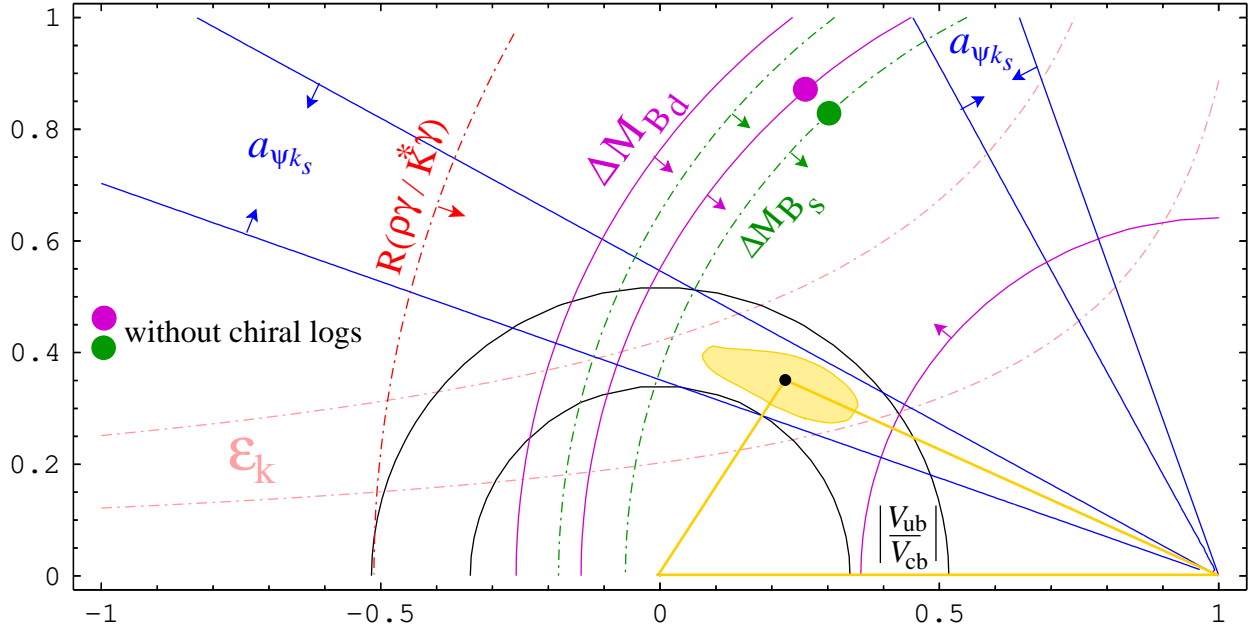}} }}
\vskip 0.2cm
\caption[1]{\it Unitary triangle fit in the SM and
the resulting 95\% C.L. contour in the $\bar \rho$ - $\bar \eta$
plane. The impact of the $R(\r\g/K^*\g) < 0.047$ constraint is
also shown.}
\label{fig:utsm}
\vskip -0.2cm
\end{figure}
\begin{figure}[!t]
 \centerline{\mbox{\epsfxsize=0.8 \linewidth \hbox{\epsfbox{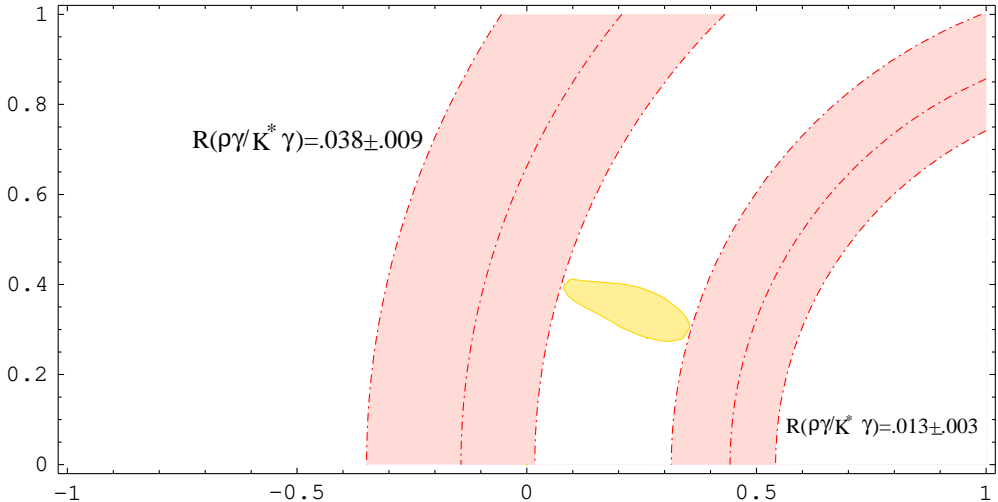}} }}
\vskip 0.2cm {
\caption[1]{ \it Extremal values of $R(\r\g/K^*\g)$ that are compatible with
the SM unitarity triangle analysis.}
\label{fig:rappsm} }
\vskip -0.2cm
\end{figure}
Note, that as in the EMFV model there are additional contributions to
the effective Wilson coefficients $C_7$, entering through the mass
insertion parameters $\delta^{u}_{13}$ (for the $b \to d \gamma$ case)
and $\delta^{u}_{23}$ (for the $b \to s \gamma$ case), which are in
general different, we have introduced two different magnetic moment WC's
for
the $d$ and $s$ sectors.  For $C_7^d = C_7^s = C_7^{\rm SM}$ we
reproduce the formulae presented in \rf{bdgAP} with the only exception
of $\D R (\r\g)$; in this case, the factor in the parentheses in the
first line of \eq{dr} is missing in \rf{bdgAP}. This, however, has
only a small numerical effect, as can be judged from the values
$\Delta R^\pm =0.055 \pm 0.130$ and $\Delta R^0=0.015 \pm 0.110$ that
we have obtained here and which are in quite good agreement with the
corresponding values given in \rf{bdgAP}.

\section{Impact on the unitarity triangle}
In this section, we present an
updated analysis of the constraints in the $(\bar \r, \bar \et)$ plane
from the unitarity of the CKM matrix, including the measurements of
the CP asymmetry $a_{\psi K_s}$ in the decays $B^0/\overline{B^0} \to
J/\psi K_s$ (and related modes), and show the impact of the upper
limit $R(\rho \gamma/K^* \gamma) \leq 0.047$~\cite{babar:convery}.

The SM expressions for $\epsilon_K$ (CP-violating parameter in $K$
decays), $\D M_{B_d}$ ($B_d^0$-$\bar B_d^0$ mass difference), $\D
M_{B_s}$ ($B_s^0$-$\bar B_s^0$ mass difference) and $a_{\psi K_s}$ are
fairly standard and can be found, for instance, in \rf{utAL}, where
also references to the various theoretical input parameters which have
not changed since then can be found.
Note that for the hadronic parameters $f_{B_d} \sqrt{\hat{B}_{B_d}}$
 and $\xi_s$, we use the recent lattice estimates
\cite{Lellouch:2002} which  into account
uncertainties induced by the so-called chiral logarithms
\cite{Kronfeld:2002ab}.
These errors are highly asymmetric and, once taken into account, reduce
sizeably the impact of the $\Delta M_{B_s}/ \Delta M_{B_d}$  lower bound
on the unitarity triangle analysis. The experimental inputs
for the quantities $\lambda$ and $\epsilon_K$
are taken from the Particle Data Group~\cite{Hagiwara:pw}. The measurement
of the CP asymmetry $a_{\psi K_s}$ in the decays
$B^0/\overline{B^0} \to J/\psi K_s$ (and related modes) is now dominated
by the BABAR~\cite{Aubert:2002ic} and BELLE~\cite{Abe:2002}
collaborations; taking into account
the earlier measurements yield the current world average $a_{\psi K_s}
=0.734 \pm 0.054$~\cite{Nir:2002gu}. The indicated value of the mass
difference $\D M_{B_d} =0.503 \pm 0.006$ ps$^{-1}$ is the current
world average~\cite{Stocchi:2002} and the 95\% C.L. lower bound $\D
M_{B_s} \geq 14.4$ ps$^{-1}$ has been recently updated this summer
~\cite{Willocq:2002cj}. The values of the theoretical parameters and
experimental
measurements that we use are summarized in Table~\ref{inputs}.

The SM fit of the unitarity triangle is presented in \fig{fig:utsm}, where
we show explicitly what happens to the allowed regions once the errors
associated with the chiral logs  are taken into account. The $\cl{95}$
contour is drawn taking into account chiral logarithm uncertainties.
The fitted values for the Wolfenstein parameters, the angles $\a$ and $\b$,
 $\D M_{B_s}$ and the CKM ratio $|V_{tb}/ V_{ub}|$ are given below where
we also show the resulting values that we obtain
without including the chiral logarithms uncertainties:
\begin{center}
\begin{tabular}{l l l} \hline
 & $\chi$-logs & no $\chi$-logs \cr \hline
$\bar\rho$ & $0.22 \pm 0.07$ & $0.25 \pm 0.07$ \cr $\bar\eta $ & $
0.34 \pm 0.04$ & $ 0.34 \pm 0.04$ \cr $ \a$ & $ (98\pm 10)^{\rm
o}$ & $(101\pm 10)^{\rm o} $ \cr $ \b $ & $(24.2 \pm 1.8)^{\rm o}
$ & $ (25.0 \pm 1.9)^{\rm o}$ \cr $ \g $ & $(60 \pm 10)^{\rm 0} $
& $ (56 \pm 10)^{\rm o}$ \cr $ \D M_{B_s}$ & $(19.6^{+4.4}_{-1.3}
)~{\rm ps}^{-1} $ & $ (21.0^{+4.8}_{-1.4}  )~{\rm ps}^{-1}$ \cr $
\left| V_{td} \over V_{ub} \right|$ & $ 1.75 \pm 0.15 $ & $ 1.61
\pm 0.14$ \cr \hline
\end{tabular}
\end{center}
The main effect of the chiral logs is that they decrease the central value
of $\bar \rho$ by about half a sigma with $\bar \eta$ remaining
practically unchanged. The largest impact of this shift is in
the increased value of the CKM matrix element ratio $ \left|
V_{td} \over V_{ub} \right|$, whose central value  moves up  by about 1
sigma.

As the bound from the current upper limit on $R(\rho \gamma/K^* \gamma)$ is
not yet competitive to the ones from either the measurement of $\Delta
M_{B_d}$, or the current bound on $\Delta M_{B_s}$, we use the allowed
$\bar \r - \bar \et$ region in order to work out the SM predictions
for the observables in the radiative $B$-decays described above.
Taking into account these errors and the uncertainties on the
theoretical parameters presented in Table~\ref{inputs}, we find the
following SM expectations for the radiative decays:
\bea
R^\pm (\r \g /K^* \g) &=& 0.023 \pm 0.012 \; ,\\
R^0 (\r \g /K^* \g) &=& 0.011\pm 0.006 \; ,\\
\D (\r \g ) &=& 0.04^{+0.14}_{-0.07} \; ,\\
A_{CP}^\pm (\r\g) &=& 0.10^{+0.03}_{-0.02} \; , \\
A_{CP}^0 (\r\g) &=& 0.06 \pm 0.02 \; .
\eea
It is interesting to work out the extremal values of $R(\rho
\gamma/K^* \gamma)$ compatible with the SM UT-analysis. This is
geometrically shown in \fig{fig:rappsm} where we draw the bands
corresponding to the values $0.038\pm 0.009$ and $0.013\pm 0.003$ (the
errors are essentially driven by the uncertainty on $\z$). The meaning
of this figure is as follows: any measurement of $R(\rho \gamma/K^*
\gamma)$, whose central value lies in the range $(0.013,0.038)$ would
be compatible with the SM, irrespective of the size of the
experimental error. The error induced by the imprecise determination
of the isospin breaking parameter $\z$ limits currently the
possibility of having a very sharp impact from $R(\rho
\gamma/K^*\gamma)$ on the UT analysis.
\begin{figure}[!t]
 \centerline{\mbox{\epsfxsize=0.8 \linewidth \hbox{\epsfbox{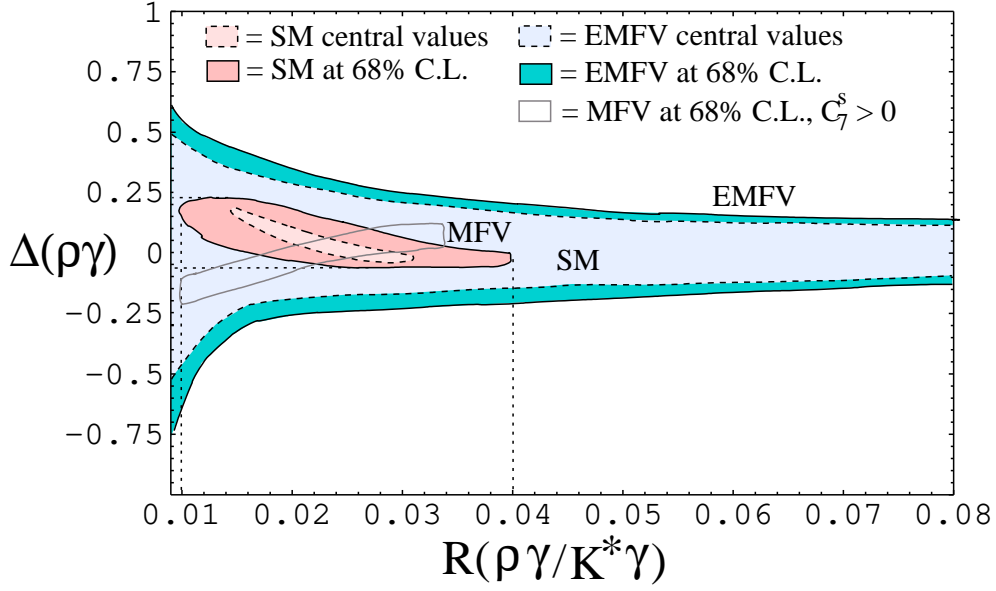}} }}
\vskip 0.2cm {
\caption[1]{ \it Correlation between $R(\r\g/K^*\g)$ and
$\Delta(\rho\gamma)$ in the SM and in MFV and EMFV models. The
light-shaded regions are obtained varying $\bar \rho$, $\bar \eta$,
the supersymmetric parameters (for the MFV and EMFV models) and using
the central values of all the hadronic quantities. The darker regions
show the effect of $\pm 1 \sigma$ variation of the hadronic
parameters.}
\label{fig:rapp_delta} }
\vskip -0.2cm
\end{figure}
\begin{figure}[!t]
\centerline{\mbox{\epsfxsize=0.8 \linewidth
\hbox{\epsfbox{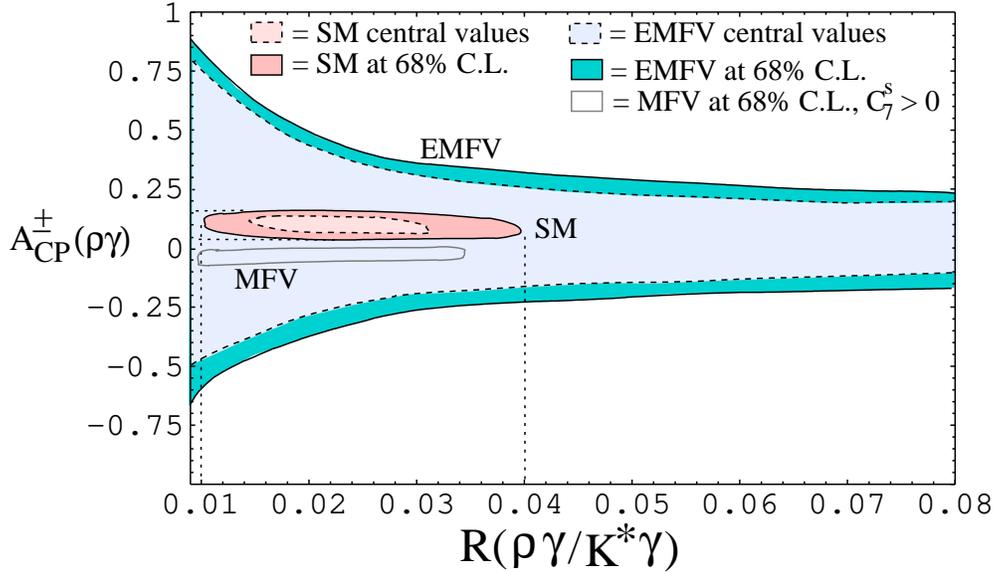}} }}
\vskip 0.2cm {
\caption[1]{ \it Correlation between $R(\r\g/K^*\g)$ and
$A_{CP}^\pm(\rho\gamma)$.
 See the caption in Fig.~\ref{fig:rapp_delta} for further details.}
\label{fig:rapp_acp} }
\vskip -0.2cm
\end{figure}

\section{Analysis in supersymmetry} We focus on two variants of the MSSM
called in the literature as MFV~\cite{mfvCDGG} and
Extended-MFV~\cite{emfvAL} models. In MFV models all the flavor
changing sources, other than the CKM matrix, are neglected and the
remaining parameters (that are assumed to be real) are the common mass
of the heavy squarks other than the lightest stop ($M_{\tilde q}$),
the mass of the lightest stop ($M_{\tilde t}$), the stop mixing angle
($\theta_{\tilde t}$), the ratio of the vacuum expectation values of
the two Higgs bosons ($\tan \b_S$), the two parameters of the chargino
mass matrix ($\m$ and $M_2$) and the charged Higgs mass
($M_{H^\pm}$). In this class of models there are essentially no
additional contributions (on top of the SM ones) to $a_{\psi K_S}$ and
$\D M_{B_s}/\D M_{B_d}$, while the impact on $\epsilon_K$, $\D
M_{B_d}$ and $\D M_{B_s}$ is described by a single parameter $f$,
whose value depends on the parameters of the supersymmetric
models~\cite{utAL}.

In EMFV models, there is an additional parameter
\beq
\d_{\tilde u_L \tilde t} = {M^2_{\tilde u_L \tilde t} \over M_{\tilde t} M_{\tilde q}}
{V_{td} \over |V_{td}|} \; .
\eeq
With the inclusion of this new parameter, the
description of the UT-related observables needs one more complex
parameter, $g = g_R + i g_I$~\cite{emfvAL}. A signature of these
models is the presence of a new phase in the $B_d^0-\bar B_d^0$ mixing
amplitude. Using the parametrization $M_{12}^d = r_d^2 e^{2 i
\theta_d} M_{12}^{\rm SM}$, we get $r_d^2 = |1+f+g|$ and $\theta_d =
1/2 \arg (1 + f + g)$. This implies supersymmetric contributions to
the CP asymmetry $a_{\psi K_s}$, which we quantify below.

\begin{figure}[!t]
 \centerline{\mbox{\epsfxsize=0.8 \linewidth  \hbox{\epsfbox{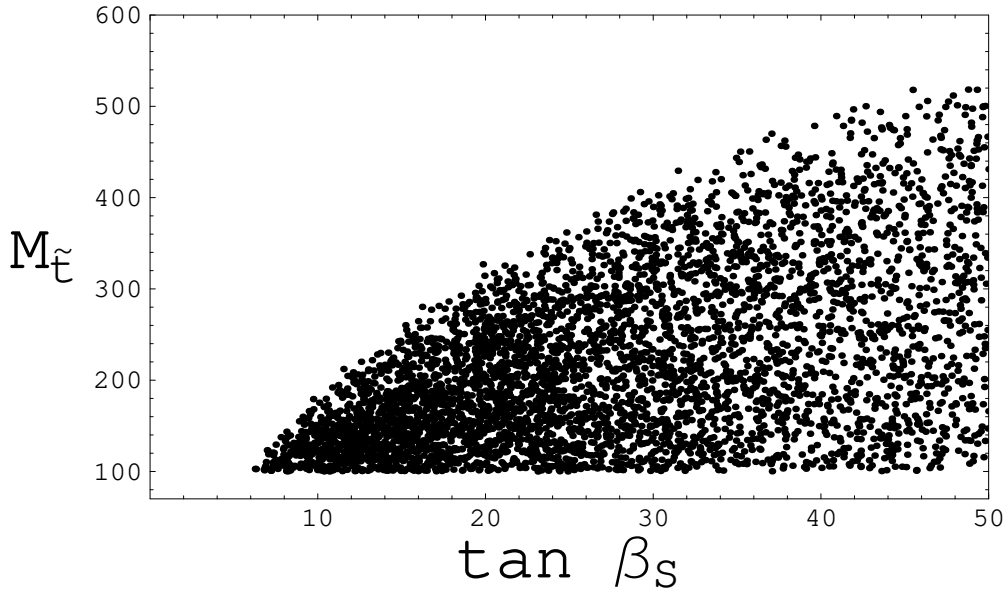}} }}
\vskip 0.2cm {
\caption[1]{ \it Correlation between $\tan \b_S$ and $M_{\tilde t}$
(in GeV) in the MFV-SUSY model for the $C_7^s>0$ scenario.}
\label{fig:tbmst} }
\vskip -0.2cm
\end{figure}

We analyze the phenomenology of the MFV and EMFV models by means of scatter plots
over the supersymmetric parameter space. In the MFV case, we scan over
the following ranges ($M_{\tilde q}$ is set to $1\; \tev$, likewise
$M_{\tilde g}$ is $O(1\;\tev)$):
\bea
M_ {\tilde t} &=& [0.1\div 1]\; \tev \; , \\
\theta_{\tilde t}&=&[-\p \div \p] \; , \\
\tan \b_S &=& [3 \div 50] \; , \\
M_2 &=& [0.1 \div 1] \; \tev\; , \\
M_{H^\pm} &=& [0.1 \div 1] \; \tev \; .
\eea
In the EMFV case, we limit the range of the
stop mixing angle to $\theta_{\tilde t}=[-0.3 \div 0.3]$ (see
discussion in \rf{emfvAL}) and add the scan over $|\d_{\tilde u_L
\tilde t}|=[0\div 1]$ and $\arg \d_{\tilde u_L \tilde t} = [-\p \div
\p]$.
\begin{figure}[!t]
\centerline{\mbox{\epsfxsize=0.8 \linewidth
\hbox{\epsfbox{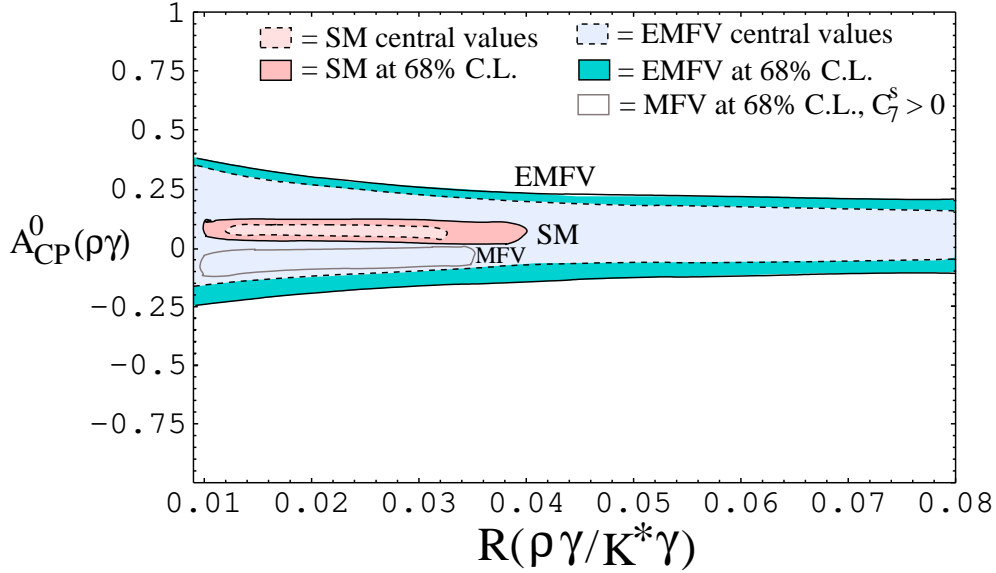}} }} \vskip 0.2cm {
\caption[1]{ \it Correlation between $R(\r\g/K^*\g)$ and $A_{CP}^0(\rho\gamma)$.
See the caption in Fig.~\ref{fig:rapp_delta} for
further details.}
\label{fig:rapp_acp0} }
\vskip -0.2cm
\end{figure}
We scan also over $\bar \rho$ and $\bar \eta$ and require that each
point satisfies the bounds that come from direct searches, from the
$B\to X_s \g$ branching ratio, and from the UT related observables
summarized in Table~\ref{inputs}. The surviving regions are presented
in Figs.~\ref{fig:rapp_delta}-\ref{fig:rapp_acp0}. In each figure, the
light shaded regions are obtained using the central values of the
input parameters given in Table~\ref{inputs} while the dark shaded
ones result from the inclusion of their one sigma errors. Note that in
the two figures showing the correlations between $A_{\rm
CP}^\pm(\rho\gamma)$ and $R(\rho\gamma/K^*\gamma)$, and $A_{\rm
CP}^0(\rho\gamma)$ and $R(\rho\gamma/K^*\gamma)$, respectively, the CP
asymmetries tend to increase as expected in the limit of small
branching ratios. In the MFV case, there are two distinct regions that
correspond to the negative (SM-like) and positive $C_7^s$ cases. For
$C_7^s<0$, the allowed regions in MFV almost coincide with the SM ones
and we do not draw them. For $C_7^s>0$, the allowed regions are
different and, in general, a change of sign of both the CP-asymmetries
(compared to the SM) is expected. We note that the latter scenario
needs very large SUSY contributions to $C_7^s$, arising from the
chargino-stop diagrams, and for fixed values of $\tan \b_S$ it is
possible to set an upper limit on the mass of the lightest stop
squark. In \fig{fig:tbmst}, we show the points that survive the $B\to
X_s \g$ constraint with a positive $C_7^s$ in the MFV scenario.  We
have also imposed the additional constraint coming from the upper
limit $\branch (B_s \to \m^+\m^-) < 2.6 \times 10^{-6}$ at
$\cl{90}$~\cite{bsmm}, and find that the allowed region is essentially
unaffected.  Note that in the $C_7^s>0$ scenario the mass of the
lightest stop has an upper bound of $500\;\gev$ for $\tan \b_S< 50$.

What concerns the allowed values of the phase $\theta_d$, comparing
the SM allowed range from the UT fit $\sin 2 \b = 0.76 \pm 0.06$,
implying $\b = 25^\circ \pm 3^\circ$, with the current experimental
value $a_{\psi K_s} = 0.734\pm 0.054$, yields $\theta_d \in
(-5^\circ,8^\circ)$. The other solution for $a_{\psi K_s}$ shown in
\fig{fig:utsm} yields $\theta_d \in (33^\circ, 46^\circ)$. The
additional contributions in $M_{12}^d$ impact on the dilepton charge
asymmetry~\cite{allRS}
$$
A_{\ell\ell} \equiv
{\ell^{++} - \ell^{--} \over \ell^{++} +\ell^{--} } =
\left(\D \G_d \over \D M_d \right)_{\rm SM} { r_d^2 \sin
2 \theta_d \over 1 + 2 r_d^2 \cos 2 \theta_d + r_d^4} \; ,
$$
where $\ell^{++}(\ell^{--})$ are the numbers of $\ell^+ \ell^+ (
\ell^- \ell^-)$ observed in the decay of a $B \bar B$ pair,
and $\D \G_d$ is the difference in the decay widhts of the two mass
eigenstate. We have
computed for each point the value of the dilepton asymmetry
$A_{\ell\ell}$ and found that the allowed range in the EMFV model is
$A_{\ell\ell} \in [-0.1,0.7] \times 10^{-2}$. This expectation has to
be compared with the current experimental bound, $A_{\ell\ell}^{\rm
exp} = (0.46 \pm 1.18 \pm 1.43) \times 10^{-2}$ \cite{sekula:2002}. We
see that the experimental precision has to be improved by one order of
magnitude in order to test the EMFV models. Note that $(\D \G_d / \D
M_d )_{\rm SM} = (1.3\pm 0.2) \times 10^{-2}$, yielding typically
$A_{\ell\ell} = O(10^{-3})$ in the SM.

\section{Summary}
We have presented here an analysis of the ratio $R(\rho
\gamma/K^* \gamma)$, involving the decays $B \to \rho \gamma$ and $B
\to K^* \gamma$, the isospin-violating asymmetry in $B \to \rho
\gamma$ decays $\Delta(\rho \gamma)$, and direct CP asymmetries
$A_{\rm CP}^\pm (\rho \gamma)$ and $A_{\rm CP}^0 (\rho \gamma)$ in the
charged and neutral $B$-meson decays in the SM and two variants of
supersymmetric theories. They illustrate the current and impending
interest in the radiative decays $B \to \rho \gamma$, which will
provide powerful constraints on the CKM parameters and allow to search
for physics beyond-the-SM.

\section*{Acknowledgments}
We would like to thank David London and
Alexander Parkhomenko for helpful correspondence and
discussions. E.L. thanks Frank Kr\"uger and Athanasios Dedes for
useful discussions, and acknowledges financial support from the
Alexander von Humboldt foundation.

\end{document}